\documentclass[prapplied,twocolumn,preprintnumbers,amsmath,amssymb,longbibliography]{revtex4-1}
\usepackage{graphicx}
\usepackage{color}      
\usepackage{dcolumn}    
\usepackage{bm}         
\usepackage[caption=false]{subfig}
\usepackage{float}
\usepackage{relsize}
\usepackage{upgreek}
\usepackage{amsmath}
\usepackage{amssymb}
\usepackage{amsthm}

\begin{document}
\title{Theory of coherent active convolved illumination for superresolution enhancement}
\author{Anindya Ghoshroy$^1$}
\author{Wyatt Adams$^1$}
\author{Durdu \"O. G\"uney$^1$}
\email{Corresponding author: dguney@mtu.edu}
\affiliation{$^1$Department of Electrical and Computer Engineering , Michigan Technological University, 1400 Townsend Dr, Houghton, MI 49931-1295, USA}

\date{\today}

\begin{abstract}
Recently an optical amplification process called the plasmon injection scheme was introduced as an effective solution to overcoming losses in metamaterials. Implementations with near-field imaging applications have indicated substantial performance enhancements even in the presence of noise. This powerful and versatile compensation technique, which has since been renamed to a more generalized active convolved illumination, offers new possibilities of improving the performance of many previously conceived metamaterial-based devices and conventional imaging systems. In this work, we present the first comprehensive mathematical breakdown of active convolved illumination for coherent imaging. Our analysis highlights the distinctive features of active convolved illumination, such as selective spectral amplification and correlations, and provides a rigorous understanding of the loss compensation process. These features are achieved by an auxiliary source coherently superimposed with the object field. The auxiliary source is designed to have three important properties. First, it is correlated with the object field. Second, it is defined over a finite spectral bandwidth. Third, it is amplified over that selected bandwidth. We derive the variance for the image spectrum and show that utilizing the auxiliary source with the above properties can significantly improve the spectral signal-to-noise ratio and resolution limit. Besides enhanced superresolution imaging, the theory can be potentially generalized to the compensation of information or photon loss in a wide variety of coherent and incoherent linear systems including those, for example, in atmospheric imaging, time-domain spectroscopy, ${\cal PT}$ symmetric non-Hermitian photonics, and even quantum computing.  
\end{abstract}
\maketitle

\section{Introduction}
Metamaterials (MMs), which are artificial inhomogeneous structures usually designed with subwavelength metal/dielectric or all-dielectric building blocks, rose to prominence nearly two decades ago as an appealing direction for designing materials with unprecedented electromagnetic properties previously considered difficult, if not impossible, to realize. Invisibility cloaks \cite{schurig2006metamaterial}, ultra-high-resolution imaging \cite{pendry2000negative,taubner2006near,jacob2006optical} and photolithography \cite{gao2015enhancing}, enhanced photovoltaics \cite{gwamuri2013advances,vora2014exchanging}, miniaturized antennas \cite{odabasi2013electrically}, ultrafast optical modulation \cite{Neira2014}, and metasurfaces \cite{arbabi2015dielectric,genevet2017recent} are few of the multitude of applications which have been envisioned. Supported by parallel efforts in micro and nanofabrication, MMs are anticipated to have broad impact on many technologies employing electromagnetic radiation. However, despite enormous theoretical and experimental progress, numerous lingering problems \cite{soukoulis2011past} require diligent consideration. Optical losses continue to be one of the greatest threats to the viability of many of the MM-based devices proposed to date. Mitigation of losses remains a challenging problem for the MM community. Gain medium was initially proposed \cite{PhysRevB.67.201101,PhysRevA.80.053807,wuestner2010overcoming,xiao2010loss} as a potential solution. However, later studies showed that stability and gain saturation issues as a result of stimulated emission near the field enhancement regions leads to intense noise generation \cite{soukoulis2010optical,PhysRevLett.98.177404,PhysRevLett.101.167401}. Due to these concerns and other associated complexities such as pump requirement, progress towards the development of a robust loss compensation scheme has been somewhat sluggish even after nearly two decades of efforts. Dielectric metasurfaces have also been proposed to alleviate some of these concerns \cite{arbabi2015dielectric,genevet2017recent}.

A recent theoretical study \cite{sadatgol2015plasmon} investigated an unconventional approach in the form of an alternative exploration of ``virtual gain'' \cite{li2019virtual,ghoshroy2020loss} to manage the losses in MMs. This compensation process, designated “plasmon injection (PI or $\Pi$) scheme,” employs an additional source to modify the field incident on a lossy MM. This auxiliary source is designed to adequately amplify an arbitrary field thereby enhancing its transmission through the MM. A multiport MM structure was used in \cite{sadatgol2015plasmon} to illustrate the conceptual operating principle for the amplification of normally incident waves. Auxiliary fields are used to coherently add energy to the lossy system to compensate the losses of different natures. This amplification mechanism was related in \cite{krasnok2020active,ghoshroy2020loss} to coherent amplification of pulses using a passive cavity described in \cite{jones2002femtosecond,potma2003picosecond}. The main difference in \cite{sadatgol2015plasmon} is continuous wave operation at the nanoscale plasmonic MM structure. In \cite{ghoshroy2017active,PhysRevApplied.10.024018,Ghoshroy:18}, we discussed in detail the generalization of the $\Pi$ scheme to imaging, which involves a spectrum of spatial frequencies. A systematic amplification in the Fourier spectrum plays a key role in extending the resolution limit of the imaging system. This is akin to the Wiener optimal filtering principle that also attempts to cleverly privilege spatial frequencies with respect to their noises \cite{roggemann1992linear,biemond1990iterative,zaknich2005principles}. The earlier theoretical studies with MM or near-field imaging systems employing negative index materials (NIMs) \cite{adams2016bringing}, superlenses \cite{Adams:17}, and hyperlenses \cite{zhang2016enhancing,Zhang:17} produced promising results. Implementing the $\Pi$ scheme with the above systems resulted in performance improvements. The distinguishing feature of the $\Pi$ scheme is the auxiliary source. The earlier variants of the $\Pi$ scheme were shown to emulate linear deconvolution \cite{adams2016bringing}.

The physical generation of the auxiliary source requires some considerations. It was shown \cite{ghoshroy2017active} that the auxiliary source can be generated through a convolution process with the original object field incident at the detector while selectively providing amplification to a controllable band of spatial frequencies. As a result of this process, the auxiliary source becomes correlated with the original object field \cite{ghoshroy2017active,qian2017emerging}. A near-field spatial filter designed with hyperbolic metamaterials (HMMs) was proposed to physically generate the auxiliary source \cite{PhysRevApplied.10.024018} with the above properties. The filter was integrated with a $50\,$nm silver film to illustrate the overall loss compensation process. This was the first potential application of the spatial filtering properties \cite{schurig2003spatial,rizza2012terahertz,liu2017nanofocusing,liang2018achieving,kieliszczyk2018tunable} of HMMs in the context of loss compensation. Later studies with coherent \cite{Ghoshroy:18} and incoherent \cite{doi:10.1021/acsphotonics.7b01242} illumination produced favourable results. An improvement in the resolution limit of a near-field silver superlens elevated the viability of the $\Pi$ scheme as an effective alternative to previously conceived loss mitigation approaches \cite{PhysRevB.67.201101,popov2006compensating,guney2009reducing,vora2014exchanging,PhysRevA.80.053807,wuestner2010overcoming,xiao2010loss} including dielectric metasurfaces \cite{arbabi2015dielectric,genevet2017recent}. Even though the techniques presented in \cite{Ghoshroy:18,doi:10.1021/acsphotonics.7b01242} possess similar properties to the original concept of the $\Pi$ scheme in \cite{sadatgol2015plasmon}, the scheme was generalized to a more encompassing term active convolved illumination (ACI) in \cite{doi:10.1021/acsphotonics.7b01242}, since it is essentially the physical convolution operation which is key to the process. Also, the $\Pi$ scheme narrows down the process to only plasmons. More recently, the ACI technique has been applied to an experimental system, where the signal-to-noise ratio (SNR) and resolution limit of a reference far-field imaging system have been significantly improved with a modest amount of amplification \cite{adams2019enhancing}. It has also been shown that the ACI offers more tolerance to pixel saturation compared to the reference system.\\
\indent In this paper, we construct a theoretical framework to provide the first comprehensive mathematical exposition of the fundamental concept of ACI for coherent illumination. Pendry's classic setup \cite{pendry2000negative} of a silver superlens operating at a wavelength of $365\,$nm is adopted since it is the simplest configuration which broadly exemplifies the rudimentary impact of optical losses such as in not only MM systems, but also different conventional and advanced linear systems. We consider the silver superlens only as a canonical example to accentuate how the ACI permits recovery of information carried by attenuated signal with minimal noise amplification. The greater scope of this paper is to develop a noise-resistant imaging theory that can be potentially generalized to a wide range of problems in various contexts related to noisy linear systems. Specific attention is drawn towards the required mechanisms, such as selective spectral amplification, physical convolution, and correlations. This study strengthens analytically, the previous results and associated assertions \cite{Ghoshroy:18,doi:10.1021/acsphotonics.7b01242} made with numerical simulations to gain physical insight into the ACI’s working principles in imaging. We conjecture that the theory of ACI can be potentially generalized to a wide variety of noisy and lossy linear systems including, for example, those in atmospheric imaging \cite{hanafy2014detailed,hanafy2015estimating,hanafy2015reconstruction,ghoshroy2019super}, time-domain spectroscopy \cite{guerboukha1018toward,ahi2019a}, optical communications \cite{gbur2002spreading,dogariu2003propagation,gbur2014partially,hyde2018controlling}, ${\cal PT}$ symmetric non-Hermitian photonics \cite{monticone2016parity,ganainy2019dawn,li2019virtual}, and even quantum computing \cite{gueddana2019can,gueddana2019toward,ghoshroy2019super}. Some discussions of how the ACI can be applied to atmospheric imaging, time-domain spectroscopy, and quantum computing can be found in \cite{ghoshroy2019super}.

\section{Near-Field Imaging System with ACI}

\begin{figure}[htbp]
\centering
{\includegraphics[height=0.25\textwidth]{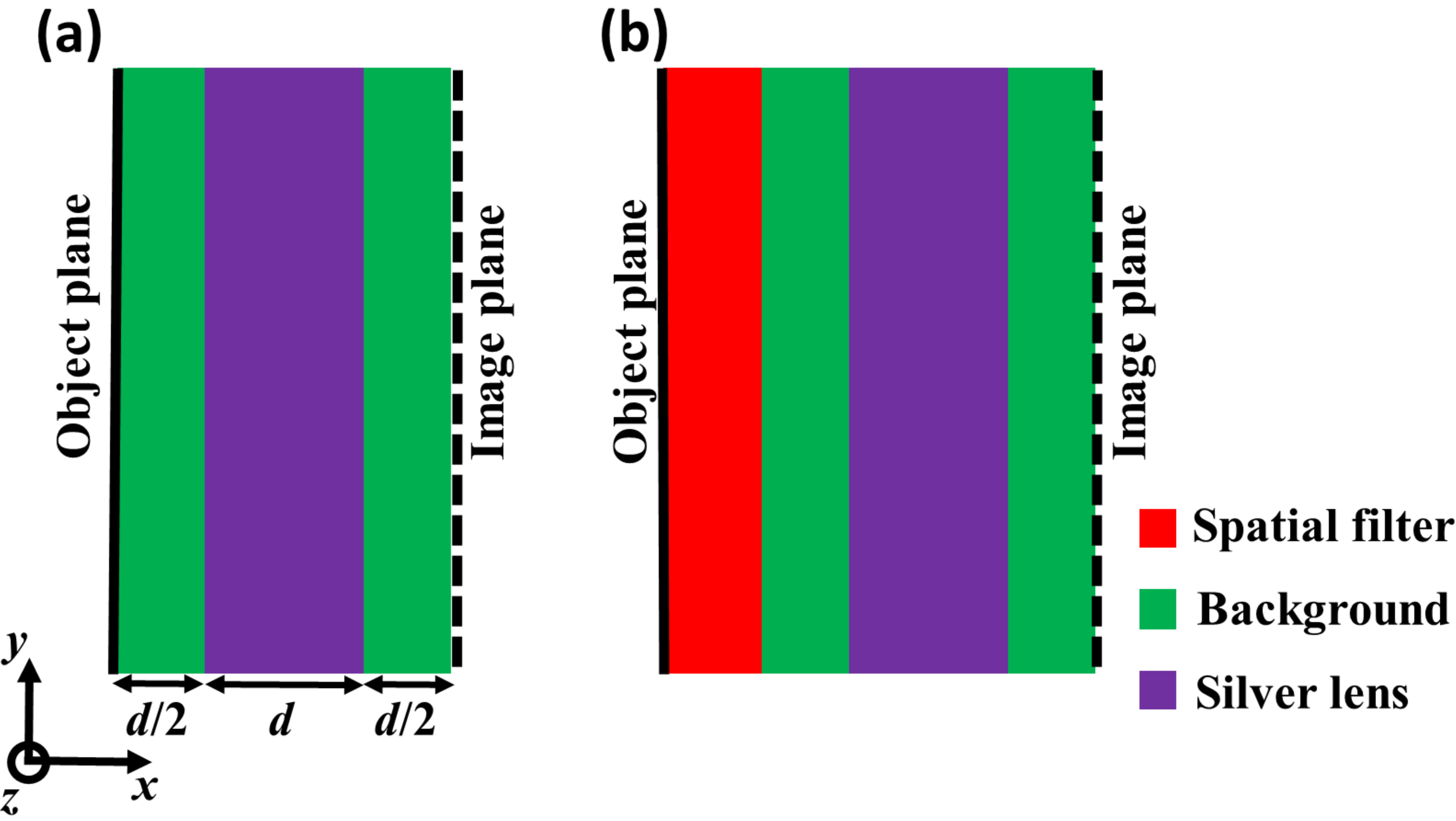}}
\caption{Schematic of a typical silver lens imaging system (a) without ACI and (b) the modified form with the integrated spatial filter for an implementation of ACI. The TM-polarized field distribution on the object plane propagates through each system and is recorded from the image plane. The purple, green, and red regions are the silver lens, background dielectric and the integrated spatial filter, respectively.}
\label{fig:Fig_Schematic}
\end{figure}

As Pendry pointed out \cite{pendry2000negative}, the properties of a NIM necessary for superresolution imaging far beyond the diffraction limit, can be attained for transverse magnetic (TM) polarized light at a wavelength $\lambda = 365\,$nm by a thin silver film embedded inside a dielectric. Under such conditions, resonant excitation of surface plasmons at the silver interface provides satisfactory amplification to high spatial frequencies which can then be focused assuming that the thickness of the silver film, object and image plane distances are much smaller than the incident wavelength. The configuration of such an imaging system is shown in Fig. \ref{fig:Fig_Schematic}(a), where the silver film with thickness $d$ is embedded inside a dielectric and positioned symmetrically between the object and image planes indicated by solid and dashed black lines, respectively. A TM field distribution on the object plane is detected from the image plane after propagating though the silver film. During this propagation process, material losses progressively degrade the transmission of high spatial frequencies with increasing transversal wavenumber $k_y$. Therefore, the ultimate performance of the system is limited to the highest spatial frequency whose attenuated amplitude is strong enough to be accurately detected from the image plane amid noise. An ideal loss compensation scheme should extend this limit by intelligently providing adequate amounts of power to these previously undetectable spatial frequencies to allow them to survive the lossy transmission process by ensuring minimal noise amplification.

In ACI, loss compensation can be performed by introducing an additional material between the object plane and the lens as shown in Fig. \ref{fig:Fig_Schematic}(b). This material should behave as a tunable active band-pass spatial filter \cite{ghoshroy2017active,Ghoshroy:18}. We write the transfer function of the spatial filter as \cite{PhysRevApplied.10.024018}
\begin{equation}
a(k_y) = b + G(k_y),
\label{eq:Passive_spatial_filter}
\end{equation}
where we set $b$ as a real constant corresponding to a uniform low background transmission. $G(k_y) = |G(k_y)|\mathrm{e}^{i \varphi(k_y)}$ is a complex band-limited function with phase $\varphi(k_y)$ and describes the pass-band of the passive filter. If the amplitude of the wave illuminating the system is increased by a factor $A_0 = b^{-1}$, the resulting transmitted spectrum is
\begin{equation}
A(k_y) = 1 + A_0G(k_y).
\label{eq:Active_spatial_filter}
\end{equation}
Eq. \ref{eq:Active_spatial_filter} is defined as the transfer function of the active spatial filter \cite{PhysRevApplied.10.024018}. The term ``active spatial filter" simply refers to the process of physically providing increased energy to the passive filter with the transfer function in Eq. \ref{eq:Passive_spatial_filter}. In other words, linear transmission through passive materials is considered. The word active also distinguishes ACI from purely deconvolution based methods \cite{zhang2016enhancing,adams2016bringing,Zhang:17,Adams:17} where no additional energy is provided to the system.

The response of the active spatial filter should be shift invariant along the object plane and integrating the filter with the lens should allow the entire system to be described with an active transfer function \cite{ghoshroy2017active,Ghoshroy:18} written as
\begin{equation}
T_{A}(k_y) = T(k_y)[1 + A_0G(k_y)],
\label{eq:Integrated_transfer_function}
\end{equation}
where $T(k_y)$ is the passive transfer function of the silver lens.
\begin{figure}[htbp]
\centering
\includegraphics[width=0.95\linewidth]{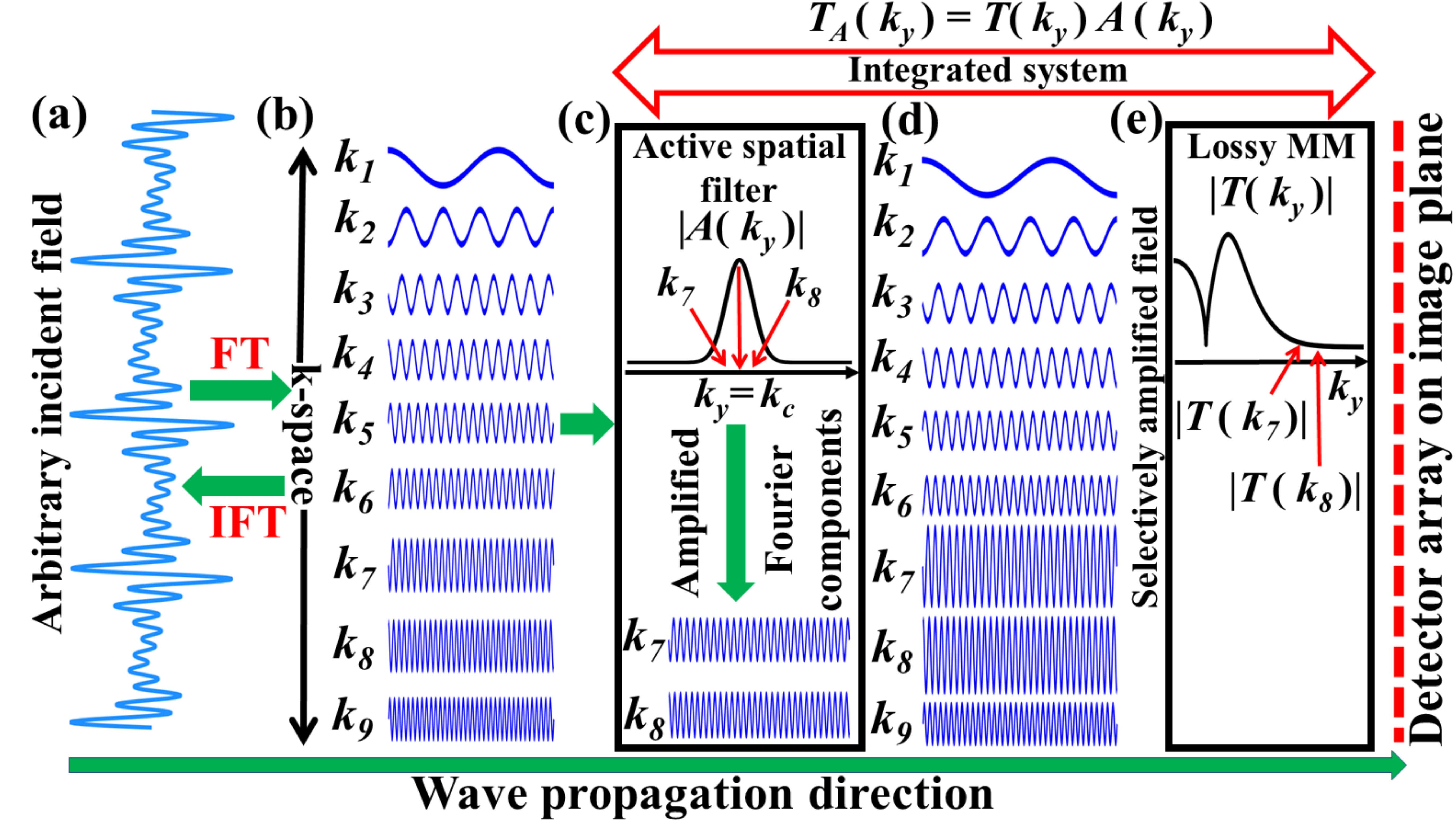}
\caption{Conceptual schematic of ACI emphasizing the underlying physics of the loss compensation process. (a) An arbitrary incident field is (b) a weighted superposition of different harmonics. After propagating though (c) the active spatial filter, the amplitudes of a few selected harmonics (e.g., $k_7$ and $k_8$) are amplified relative to the others based on the transfer function of the filter given by Eq. \ref{eq:Active_spatial_filter} and depicted by the inset in (c). (d) The modified field incident on (e) the lossy MM contain the amplified spatial frequencies (origin of auxiliary source), superimposed with the unaltered harmonics. The estimated amplification provided to the selected harmonics can be the inverse of their passive transmission amplitudes (e.g., $T(k_7)$ and $T(k_8)$ in the inset). This would ensure that they survive the lossy transmission process through the MM.}
\label{fig:Fig_Schematic2}
\end{figure}

The Eqs. \ref{eq:Active_spatial_filter} and \ref{eq:Integrated_transfer_function} are central to the ACI loss compensation process. The physical picture is best illustrated with the aid of the schematic shown in Fig. \ref{fig:Fig_Schematic2}. An arbitrary field incident on the system, shown in Fig. \ref{fig:Fig_Schematic2}(a) can be described as a linear, weighted superposition of harmonics or spatial frequencies shown in Fig. \ref{fig:Fig_Schematic2}(b). The active spatial filter in Fig. \ref{fig:Fig_Schematic2}(c) is inserted between the lossy MM in Fig. \ref{fig:Fig_Schematic2}(e) and the incident field. The transfer function of the active filter has the form of Eq. \ref{eq:Active_spatial_filter} and the amplitude $|A(k_y)|$ is depicted in the inset in Fig. \ref{fig:Fig_Schematic2}(c). The transmission amplitude of a lossy MM, which deteriorates for increasing spatial frequencies, is illustrated by the inset in Fig. \ref{fig:Fig_Schematic2}(e). For example, assume that the spatial frequencies $k_7$ and $k_8$ will be compensated. ACI achieves this by tuning the center frequency, $k_c$ of the active spatial filter such that $|A(k_y)|>1$ over the identified spatial frequencies. This is illustrated by the inset in Fig. \ref{fig:Fig_Schematic2}(c). After the harmonics of the incident field propagate though the active spatial filter, the amplitudes of the identified spatial frequencies are amplified relative to the other harmonics. Therefore, the field exiting the active filter contains the original harmonics of the object superimposed with the selectively amplified harmonics $k_7$ and $k_8$ [see Fig. \ref{fig:Fig_Schematic2}(d)]. The amplification provided to these harmonics (controlled by $A_0$) is adjusted to ensure that they survive the lossy transmission process through the MM. The selectively amplified spatial frequencies at the exit of the filter constitute the auxiliary source as discussed in \cite{ghoshroy2017active} and is conceptually similar to \cite{sadatgol2015plasmon} with the only difference being the generation process, which here like in \cite{ghoshroy2017active} employs the active spatial filter to simply modify the original field incident on the MM by a convolution operation.

Using numerical simulations, ACI was implemented with an experimentally realized silver superlens \cite{fang2005sub} at the wavelength $\lambda = 365\,$nm. A physical system approximating the properties of the active spatial filter was designed with aluminium-dielectric multilayered structures which exhibit hyperbolic dispersion \cite{PhysRevApplied.10.024018}.  The above theoretical formulation was tested by integrating the multilayered structure with the lens as shown in Fig. \ref{fig:Fig_Schematic}(b). Imaging results with coherent \cite{Ghoshroy:18} and incoherent \cite{doi:10.1021/acsphotonics.7b01242} illumination showed an improvement in the resolution limit of the lens even under the presence of noise. The HMMs used in \cite{Ghoshroy:18,doi:10.1021/acsphotonics.7b01242} were designed to act as the spatial filter in Fig. \ref{fig:Fig_Schematic}(b), such that their transmission properties closely approximate Eq. \ref{eq:Active_spatial_filter} under high intensity illumination.

\section{Variance in the Fourier Domain} 

\begin{figure}[htbp]
\centering
\includegraphics[width=0.95\linewidth]{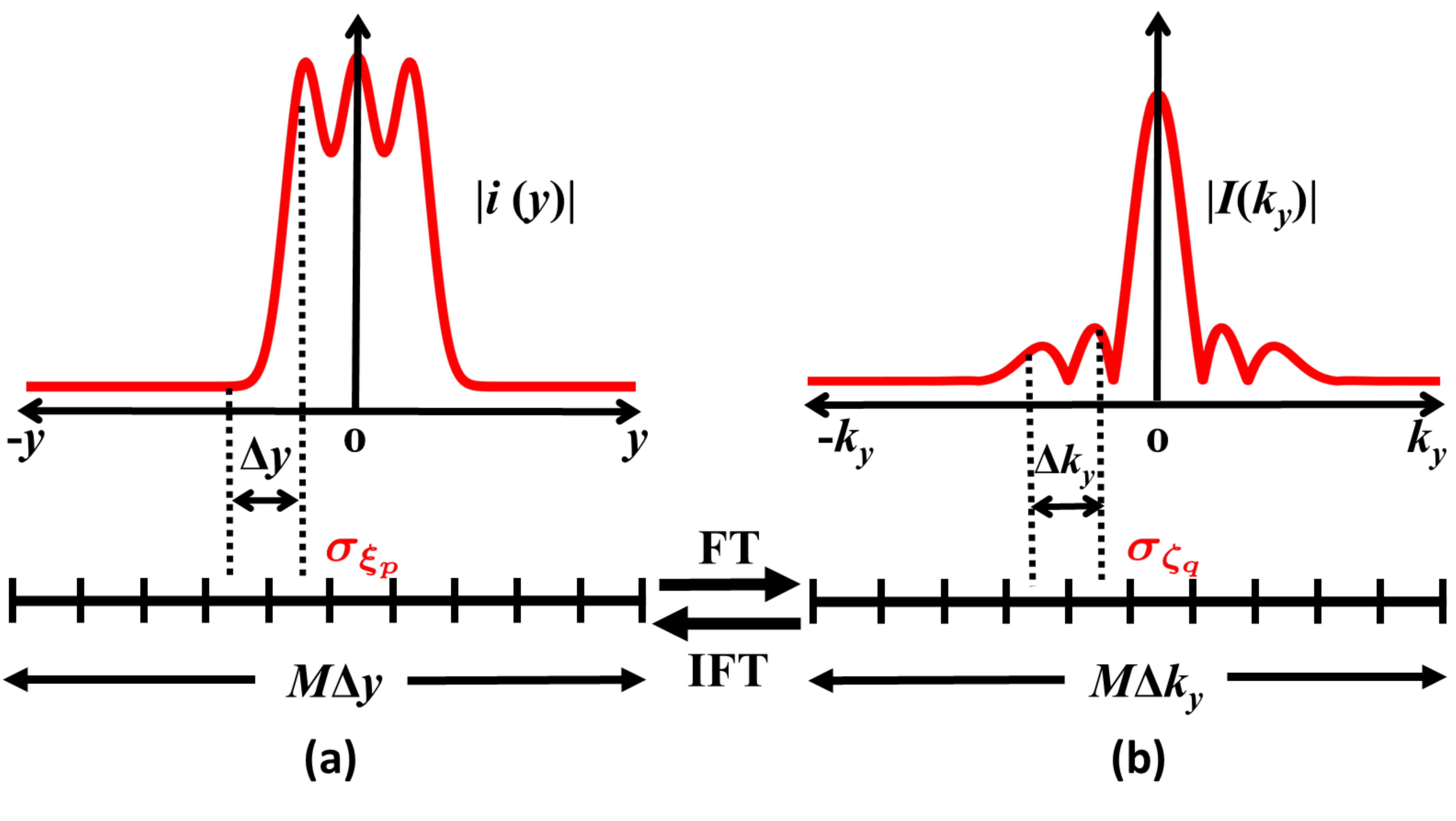}
\caption{Illustration of an image measurement process in a detector system. (a) The continuous field $i(y)$ along the image plane of length $L$ is decomposed into $M$ samples at intervals of $\Delta y$. Each sample is identified by an integer $p$. During the detection process, noise degrades the ideal image and the standard deviation at each sample is $\sigma_{\xi _p}$.$|i_{n}(\xi)|$ is the magnitude of the recorded image at the $p^{\mathrm{th}}$ pixel or $y = p\Delta y \equiv \xi$ spatial coordinate. (b) The Fourier spectrum of $i(\xi)$ is similarly a decomposition into $M$ spatial frequencies and $\sigma_{\zeta _q}$ is the standard deviation at the $q^{\mathrm{th}}$ frequency.}
\label{fig:Fig_Schematic3}
\end{figure}

To start with, let the image plane has a length $L$ along the y-axis [see Fig. \ref{fig:Fig_Schematic3}(a)]. A continuous signal $i(y)$, along the image planes is measured by a detector which can be an array of pixels or a scanning near-field probe. Based on the setup in Fig. \ref{fig:Fig_Schematic}, $i(y)$ represents TM-polarized field. An arbitrary spatial field distribution is decomposed into $M$ discrete samples at intervals of $\Delta y$ where $M$ is an even integer. The above spatial decomposition is  represented by the segmented line in Fig. \ref{fig:Fig_Schematic3}(a) where each segment is defined as a pixel. An integer $p$ satisfying $-\frac{M}{2} \leq p \leq \frac{M}{2}-1$ uniquely identifies each pixel centered at $y(p) = p\Delta y \equiv \xi$. This relates the discrete space $\xi$ to continuous space $y$. The signal sampled by the $p^{\mathrm{th}}$ pixel is denoted by $i(\xi)$. In subsequent calculations we will set $L = 80\lambda$ with $\lambda=365\,$nm and $M=5840$. Therefore, the sampling interval is $\Delta y = 5\,$nm which is slightly larger than previously demonstrated apertureless probes which can achieve resolutions down to $1\,$nm \cite{ZenhausernSNOM}. The resulting noisy image at each pixel is described with a popular signal-modulated noise model \cite{Heine:06,walkup1974image,Kasturi:83,Froehlich:81,Sadhar}. The image plane is thought of as an array of statistically independent random variables (RVs) and the subsequent noisy image at the $p^{\mathrm{th}}$ pixel is denoted by
\begin{equation}
i_{n}(\xi) = \bigg [|i(\xi)| + n_{sd}(\xi)\bigg ]\mathrm{e}^{i\theta(\xi)}.
\label{eq:Signal_modulated_noise_model}
\end{equation}
$i(\xi)$ is the noiseless field at the image plane and is corrupted by signal-dependent (SD) noise process $n_{sd}(\xi)$. The discussion of signal-independent noise can be found in \cite{ghoshroy2017active,PhysRevApplied.10.024018}. The RV $n_{sd}(\xi)$ in Eq. \ref{eq:Signal_modulated_noise_model} has zero mean, Gaussian probability density function, and standard deviation $f\{ |i(\xi)|\}^ {\gamma}\sigma _{sd}$, where $\sigma _{sd}$ is a constant. $\theta(\xi)$ is the phase of the noiseless coherent field $i(\xi)$ at the $p^{\mathrm{th}}$ pixel (i.e., $\xi \equiv p \Delta y$). A correction due to the shift in the zero-optical-path difference point based on an interferometric setup was not included in our model. $f\{|i(\xi)|\}^ \gamma$ is a function of the ideal image amplitude and is referred to as the modulation function \cite{Heine:06}. $\gamma$ is a parameter satisfying $0 \leq \gamma \leq 1$ \cite{walkup1974image}. The variance at each pixel is,
\begin{equation}
\sigma_{\xi _p} ^2 = f\{ |i(\xi)|\}^ {2\gamma} \sigma _{sd}^2.
\label{eq:Total_sigma}
\end{equation}
The modulation function and the value of $\gamma$ are selected to best mimic the behavior of SD noise which affects the system.

The above detection process results in a similar decomposition of the continuous Fourier spectrum of $i(y)$ into $M$ discrete spatial frequencies as illustrated by the segmented line in Fig. \ref{fig:Fig_Schematic3}(b). Two adjacent frequencies are separated by $\Delta k_y$ and the individual spatial frequencies are referenced by $k_y(q) = q\Delta k_y \equiv \zeta$, where $-\frac{M}{2} \leq q \leq \frac{M}{2} -1$. This relates the discrete Fourier space $\zeta$ to continuous Fourier space $k_y$. The Fourier transform of the discretized noise-free image, $i(\xi)$, is denoted by $I(\zeta)$, and the standard deviation at the $q^{\mathrm{th}}$ spatial frequency is $\sigma_{\zeta _q}$. Knowledge of $\sigma_{\zeta _q}$ is particularly useful in determining the maximum achievable limiting resolution for optical systems where transmission progressively worsens for high spatial frequencies. For example, the Fourier components with transmitted amplitudes comparable to, or less than $\sigma_{\zeta _q}$ will be indiscernible from random noise fluctuations within the measured signal. Therefore, $\sigma_{\zeta _q}$ allows us to identify the spatial frequencies whose Fourier domain information is effectively lost due to noise effects. Additionally, the effectiveness of a loss compensation technique can also be evaluated by monitoring its effect on $\sigma_{\zeta _q}$. Thus, a formulation of $\sigma_{\zeta _q}$ is important for our understanding of the underlying mechanism of ACI and its capacity at compensating losses while minimizing noise amplification.
A general expression for the standard deviation at the $q^{\mathrm{th}}$ spatial frequency can be calculated by approximating the analytical Fourier transform relation as a Riemann sum \cite{Voelz}. The Fourier transform of the noisy image $i_{n}(\xi)$ in Eq. \ref{eq:Signal_modulated_noise_model} is then written as
\begin{align}
I_{n}(\zeta)  &= \mathlarger{\sum \limits ^{(\frac{M}{2}-1)\Delta y} _{\xi=-\frac{M}{2}\Delta y}} |i_{n}(\xi)|\mathrm{e}^{i\theta(\xi)} \exp\bigg(-{i \xi \zeta}\bigg) \Delta y \notag \\
          &= \mathlarger{\sum \limits ^{(\frac{M}{2}-1)\Delta y} _{\xi=-\frac{M}{2}\Delta y}} |i_{n}(\xi)| \bigg \{ \cos \bigg(\theta(\xi) - { \xi \zeta} \bigg) \notag  \\
         &  +i \sin\bigg(\theta(\xi)-{ \xi \zeta} \bigg)  \bigg \} \Delta y \notag \\
         & = I^\prime_{n}(\zeta) +i I^{\prime \prime}_{n}(\zeta)
\label{eq:DFT_equation}%
\end{align}
with real and imaginary parts  $I^\prime_{n}(\zeta)$ and $I^{\prime \prime}_{n}(\zeta)$, respectively. Note that the number of samples $M$, is related to $\Delta y$ and $\Delta k_y$ by $M=2 \pi/\Delta y \Delta k_y$ \cite{Voelz}. We can substitute $|i_{n}(\xi)|$ from Eq. \ref{eq:Signal_modulated_noise_model} into Eq. \ref{eq:DFT_equation} to express the real and imaginary parts of $I_{n}(\zeta)$ as
\begin{equation}
I^\prime_{n}(\zeta) = \mathlarger{\sum \limits ^{(\frac{M}{2}-1)\Delta y} _{\xi=-\frac{M}{2}\Delta y}} \bigg \{ |i(\xi)| + n_{sd}(\xi) \bigg \} \cos[\phi(\xi,\zeta)] \Delta y,
\label{eq:DFT_equation_real part}
\end{equation}
and
\begin{equation}
I^{\prime \prime} _{n}(\zeta) = \mathlarger{\sum \limits ^{(\frac{M}{2}-1)\Delta y} _{\xi=-\frac{M}{2}\Delta y}} \bigg \{ |i(\xi)| + n_{sd}(\xi) \bigg \}\sin [\phi(\xi,\zeta)] \Delta y,
\label{eq:DFT_equation_imag part}
\end{equation}
respectively, and $\phi(\xi,\zeta) = {\theta(\xi)-\xi \zeta}$.

Based on Eq. \ref{eq:DFT_equation}, we can write $I_{n}(\zeta)$ as
\begin{equation}
I_{n}(\zeta)  = I(\zeta) + N_{sd}(\zeta),
\label{eq:Transformed_Noisy_Image}
\end{equation}
where $I(\zeta)$ and $N_{sd}(\zeta)$ are the Fourier transforms of $i(\xi)$ and $n_{sd}(\xi)$ in Eq. \ref{eq:Signal_modulated_noise_model}, respectively. The real and imaginary parts of $I(\zeta)$ and $N_{sd}(\zeta)$ can also be expressed in terms of the sums of cosines and sines similar to $I_{n}(\zeta)$ (see Eqs. \ref{eq:DFT_equation_real part} and \ref{eq:DFT_equation_imag part}). $N_{sd}(\zeta)$  in Eq. \ref{eq:Transformed_Noisy_Image} has a standard deviation $\sigma_{\zeta _q}$ describing SD noise at the $q^{\mathrm{th}}$ Fourier component. The variance of the real and imaginary parts of $N_{sd}(\zeta)$ are denoted by $\sigma^{2}_{\zeta _q,r}$ and $\sigma^{2} _{\zeta _q,i}$, respectively. According to Eqs.  \ref{eq:DFT_equation} and \ref{eq:Transformed_Noisy_Image}  $N_{sd}(\zeta)$ is a weighted superposition of all the RVs in the spatial domain. Each RV involved in the summation is statistically independent with a Gaussian probability density function. Therefore, we can apply Bienaym\'e's identity, to express $\sigma^{2}_{\zeta _q,r}$ and $\sigma^{2}_{\zeta _q,i}$ as
\begin{equation}
\sigma^{2} _{\zeta _q,r} = \mathlarger{\sum \limits ^{(\frac{M}{2}-1)\Delta y} _{\xi=-\frac{M}{2}\Delta y}}   \sigma^{2} _{\xi _p} \cos ^2 [\phi(\xi,\zeta)](\Delta y)^2,
\label{eq:Var_real}
\end{equation}
and
\begin{equation}
\sigma^{2} _{\zeta _q,i}= \mathlarger{\sum \limits ^{(\frac{M}{2}-1)\Delta y} _{\xi=-\frac{M}{2}\Delta y}}   \sigma^{2} _{\xi _p} \sin ^2 [\phi(\xi,\zeta)](\Delta y)^2,
\label{eq:Var_imag}
\end{equation}
respectively, and $\sigma^{2} _{\xi _p} = f\{ |i(\xi)|\}^ {2\gamma}\sigma^2 _{sd} $. The overall variance at each spatial frequency is simply the sum of the variances of the real and imaginary parts in Eqs. \ref{eq:Var_real} and \ref{eq:Var_imag}, that is
\begin{equation}
\sigma_{\zeta _q}^2  = \mathlarger{\sum \limits ^{(\frac{M}{2}-1)\Delta y} _{\xi=-\frac{M}{2}\Delta y}} f\{ |i(\xi)| \}^{2\gamma} \sigma _{sd}^2 (\Delta y)^2.
\label{eq:Sigma_final}%
\end{equation}

Without loss of generality, the subsequent calculations can be simplified and provide more physical insight by assuming the modulation function in Eq. \ref{eq:Sigma_final} is a linear function of $|i(\xi)|$  with $\gamma = 1$. This results in constant SNR and has implications on the considered noise levels and required illumination intensities. This is chosen to relate the variance directly to the total physical power contained in the signal as shown below. Similar effects are obtained, such as in practical detectors with the Poisson distribution of photon noise \cite{doi:10.1021/acsphotonics.7b01242,Heine:06,adams2019enhancing}. Substituting $f\{ |i(\xi)| \} = |i(\xi)|$ and $\gamma = 1$ we can rewrite Eq. \ref{eq:Sigma_final} as
\begin{equation}
\sigma_{\zeta _q}^2  = \bigg [ \mathlarger{\sum \limits ^{(\frac{M}{2}-1)\Delta y} _{\xi=-\frac{M}{2}\Delta y}} |i(\xi)|^2\sigma _{sd}^2 \Delta y \bigg ]\Delta y.
\label{eq:Sigma_final1}%
\end{equation}
The summation enclosed inside brackets, is proportional to the optical power on the image plane. Therefore, we can employ the energy conservation theorem by using Parseval's relation and rewrite $\sigma_{\zeta _q}^2$ in Eq. \ref{eq:Sigma_final1} as
\begin{align}
\sigma_{\zeta _q}^2 &= \bigg [\frac{1}{2\pi} \mathlarger{\sum \limits ^{(\frac{M}{2}-1)\Delta k_y} _{\zeta=-\frac{M}{2}\Delta k_y}}  {|I(\zeta)|}^{2} \sigma _{sd}^2 \Delta k_y \bigg ] \Delta y \notag \\
             &= \frac{1}{M}\mathlarger{\sum \limits ^{(\frac{M}{2}-1)\Delta k_y} _{\zeta=-\frac{M}{2}\Delta k_y}}  {|I(\zeta)|}^{2} \sigma _{sd}^2.
\label{eq:Sigma_final_fourier}
\end{align}
Eqs. \ref{eq:Sigma_final1} and \ref{eq:Sigma_final_fourier} state, for a fixed number of pixels $M=2\pi/\Delta y \Delta k_y$, that the spectral variance $\sigma_{\zeta _q}^2$ is constant and proportional to the total power contained in the signal. Similar results have been reported for incoherent light in \cite{ingerman2018signal,becker2018better,lucke2001fourier}. This is a remarkable result, which can be potentially generalized to a wide variety of problems in noisy and lossy linear systems, either classical or quantum. Below, in the context of superresolution imaging, we demonstrate how this result leads to enhanced spectral SNR with the incorporation of selective spectral amplification and correlations. For different values of $\gamma$, the variance is still flat, but not proportional to the power contained in the signal (see Eq. \ref{eq:Sigma_final}). To the best of our knowledge, the utilization of Eqs. \ref{eq:Sigma_final1} and \ref{eq:Sigma_final_fourier} in imaging has only been drawn attention to here and in a slightly modified form recently in \cite{becker2018better} to extend the SNR limit using sub-pupils.

The presence of an extra $\Delta y$ clearly makes $\sigma_{\zeta _q}^2$ dependent on the spatial discretization. Rescaling $\Delta y$ in  Eq. \ref{eq:Sigma_final_fourier} would result in effects of upsampling or downsampling of continuous signals. Therefore, Eq. \ref{eq:Sigma_final_fourier} cannot be readily generalized for an arbitrary detector system without considering the physical mechanism through which information is extracted. The spectral variance may not necessarily reduce with pixel miniaturization and multiple factors must also be considered when determining the overall effect on noise. The number of detected photons are also intimately related to the pixel active area, quantum efficiency, the pixel optical path, integration time, and sensitivity \cite{Agranov,Bigas2006,Shcherback2003,mitrofanov2001collection}. Additionally, it may be necessary to incorporate crosstalk effects between adjacent pixels to accurately model the effect of pixel scaling on $\sigma_{\zeta _q}^2 $. However, the effects of pixel miniaturization on the detected noise are considered independent from ACI, which only deals with compensation of signal losses for a fixed number of pixels.

\begin{figure}[htbp]
\centering
\includegraphics[width=0.95\linewidth]{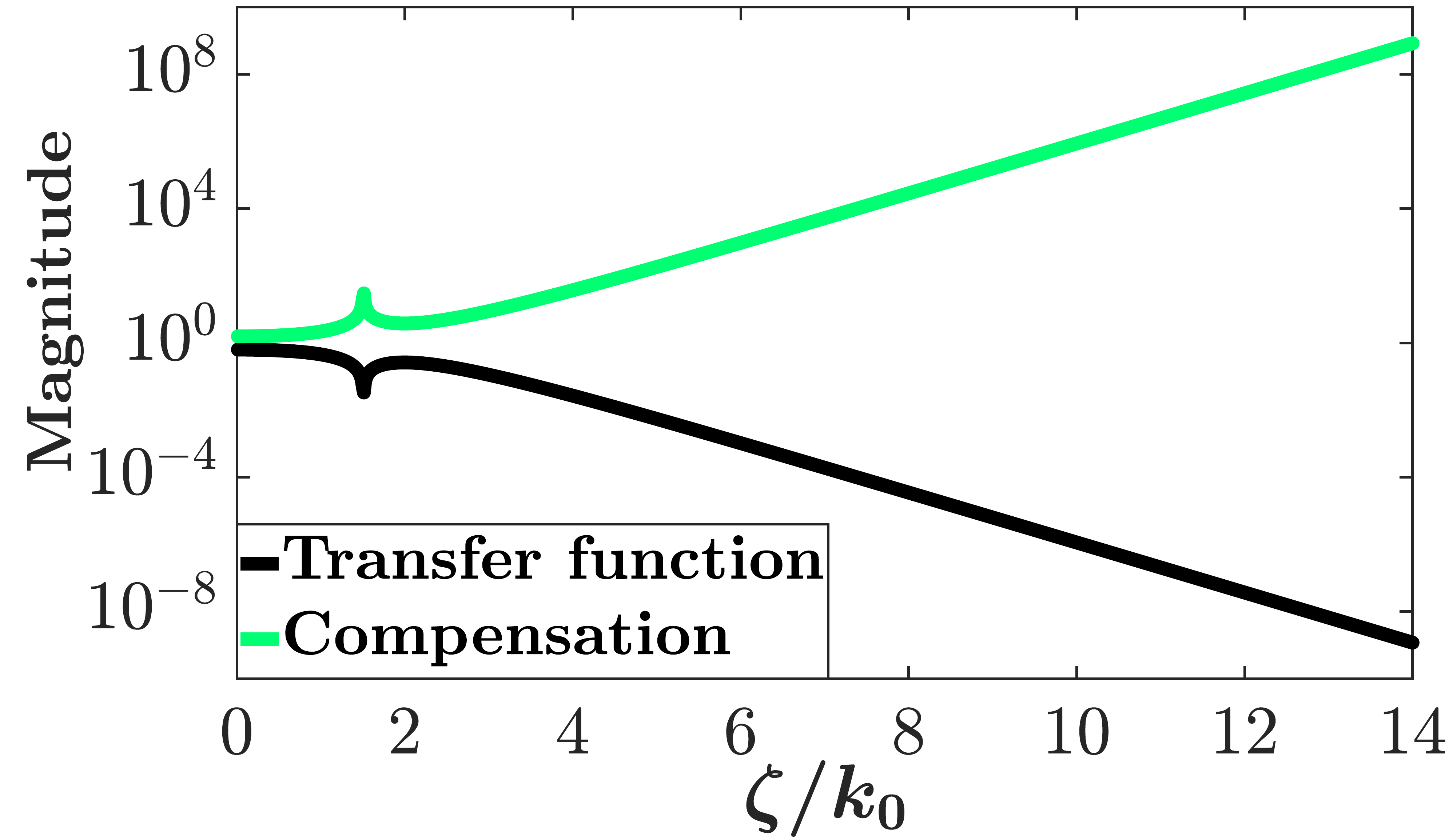}
\caption{Magnitude of the analytical transfer function for a $50\,$nm thick silver lens embedded inside a dielectric and symmetrically placed between the object and image planes as shown in Fig. \ref{fig:Fig_Schematic}(a). The required compensation at each Fourier component should ideally be the inverse of the transfer function and is shown by the green line. }
\label{fig:Fig_TransferFunction}
\end{figure}

In subsequent discussions, an analytical equation \cite{Fang:03} is used for the transfer function of the silver lens imaging system, which is configured similar to an experimental silver lens \cite{fang2005sub} with $d=50\,$nm and embedded inside a background dielectric of relative permittivity $\epsilon _d = 2.5$ \cite{PhysRevApplied.10.024018,Ghoshroy:18,doi:10.1021/acsphotonics.7b01242} [see Figs. \ref{fig:Fig_Schematic}(a) and \ref{fig:Fig_TransferFunction}]. The relative permittivity of silver at $\lambda = 365\,$nm is $\epsilon _{Ag} = -1.88 +0.60i$, calculated from the Drude-Lorentz model \cite{rakic1998optical}. The corresponding estimated compensation necessary for each spatial frequency is simply the inverse of the corresponding magnitude of transmission (see Fig. \ref{fig:Fig_TransferFunction}). In Fig. \ref{fig:Fig_TransferFunction} and following ones negative spatial frequencies are not shown, although the full spectrum is considered in all the calculations. 

\section{Results}

\subsection{Selective spectral amplification and correlations} 
\begin{figure}[htbp]
\centering
\includegraphics[width=0.95\linewidth]{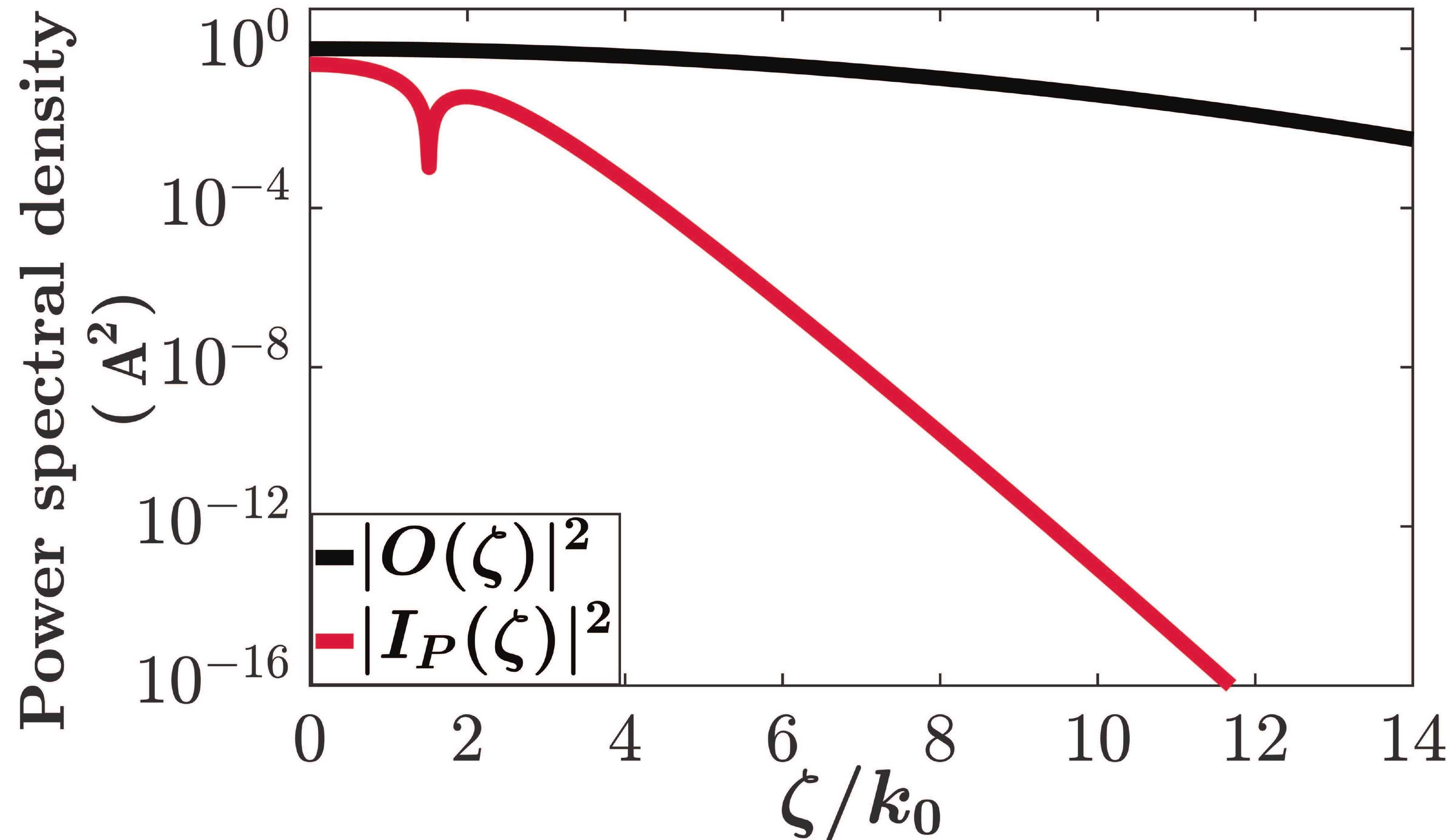}
\caption{The power spectral densities of the object ($|O(\zeta)|^2$) and the passive image ($|I_P(\zeta)|^2$) showing how the total power contained within the Gaussian object and image is distributed throughout the Fourier spectrum.}
\label{fig:Fig_PSD}
\end{figure}

In the following, an example object with a Gaussian spectrum is employed in Figs. \ref{fig:Fig_PSD} and \ref{fig:Fig_SNR}, defined as
\begin{equation}
|O(\zeta)| = \exp \bigg[ \frac{-\zeta^2}{2\alpha^2} \bigg ],
\label{eq:Object_spectrum}
\end{equation}
where $\alpha$ describes the full width at half maximum (FWHM) of $|O(\zeta)|$ and is defined as 
\begin{equation}
\alpha = \frac{0.25M\Delta k_y}{2\sqrt{2\ln 2}}.
\label{eq:Object_alpha}
\end{equation}

The imaging systems are illuminated with a TM-polarized source from the object plane (see Fig. \ref{fig:Fig_Schematic}). We assume that the object field is created through subwavelength slits \cite{taubner2006near,liu2007far}. The spatially coherent discretized complex magnetic field distribution along the object plane is denoted as $o(\xi)$. The Fourier transforms of the noiseless image for the passive (i.e., without ACI) and active (i.e., with ACI) imaging systems are
\begin{subequations}
\begin{align}
& I_{P}(\zeta) = O(\zeta)T(\zeta),               \label{eq:Images_coherent_passive}\\
& I_{A}(\zeta) = O(\zeta)T(\zeta)[1 + A_0G(\zeta)],  \label{eq:Images_coherent_active}
\end{align}
\label{eq:Images_coherent}%
\end{subequations}
respectively, where $O(\zeta) = \mathcal{F}\{o(\xi)\}$ and $\mathcal{F}$ is the Fourier transform operator. The subscripts ``$P$" and ``$A$" refer to the passive and active imaging systems, respectively. $O(\zeta)A_0G(\zeta)$ in Eq. \ref{eq:Images_coherent_active} is defined as the auxiliary source \cite{ghoshroy2017active,PhysRevApplied.10.024018,Ghoshroy:18} (see Fig. \ref{fig:Fig_Schematic2}). Therefore, $O(\zeta)T(\zeta)A_0G(\zeta)$ is the residual auxiliary source which survived the lossy transmission process through the lens. As can be seen from Eq. \ref{eq:Images_coherent_active}, the object field $O(\zeta)$ is superimposed coherently with the auxiliary source. The auxiliary source is required to possess three important properties. First, it is correlated with the object field $O(\zeta)$ \cite{ghoshroy2017active}. Second, it is defined over a finite bandwidth through $G(\zeta)$. Third, it is amplified by a factor of $A_0$. Below, without loss of generality, we use a band-limited unit magnitude rectangular function for $G(\zeta)$. Then, overall, the auxiliary source corresponds to only a portion of the object spectrum, which is selectively amplified. However, in general, the function $G(\zeta)$ can have an arbitrary profile with a finite bandwidth \cite{PhysRevApplied.10.024018}. In this case, the auxiliary source spectrum is modified in accordance with the given function $G(\zeta)$, while being still selectively amplified and correlated with the object. In \cite{PhysRevApplied.10.024018} and \cite{Ghoshroy:18}, we show how to construct such an auxiliary source using HMMs acting as near-field spatial filters.

The standard deviations at the $q^{\mathrm{th}}$ Fourier component corresponding to $I_{P}(\zeta)$ and $I_{A}(\zeta)$ in Eq. \ref{eq:Images_coherent} are denoted by ${\sigma_{\zeta _q,P}}$ and ${\sigma_{\zeta _q,A}}$, respectively. Their expressions are determined by substituting $|I(\zeta)|$ in Eq. \ref{eq:Sigma_final_fourier} with $|I_{P}(\zeta)|$ and $|I_{A}(\zeta)|$, respectively. That is
\begin{equation}
\sigma^2_{\zeta _q,P} = \frac{1}{M}\mathlarger{\sum \limits ^{(\frac{M}{2}-1)\Delta k_y} _{\zeta=-\frac{M}{2}\Delta k_y}}  |O(\zeta)T(\zeta)|^2 \sigma _{sd}^2 ,
\label{eq:Sigma_fourier_passive_coherent}
\end{equation}
and
\begin{align}
\sigma^2_{\zeta _q,A} &= \frac{1}{M}\mathlarger{\sum \limits ^{(\frac{M}{2}-1)\Delta k_y} _{\zeta=-\frac{M}{2}\Delta k_y}} |O(\zeta)T(\zeta) |^2 \sigma _{sd}^2  |1 + A_0G(\zeta) |^2 \notag \\
               &= \sigma^2_{\zeta _q,P} + \sigma^2_{\zeta _q,Aux}.
\label{eq:Sigma_fourier_active_coherent}
\end{align}
Note that $\sigma^2_{\zeta _q,A}$ can be split into its contributing parts. $\sigma^2_{\zeta _q,Aux}$ describes the contribution to the SD noise from the residual auxiliary source and is given by
\begin{align}
\sigma^2_{\zeta _q,Aux} &= \frac{1}{M}\mathlarger{\sum \limits ^{(\frac{M}{2}-1)\Delta k_y} _{\zeta=-\frac{M}{2}\Delta k_y}}  |O(\zeta)T(\zeta)  |^{2} A_0^2\sigma _{sd}^2 \notag \\
& \times \bigg [ \frac{2G^\prime (\zeta)}{A_0} +  |G(\zeta)|^2 \bigg ],
\label{eq:Sigma_fourier_aux_coherent}
\end{align}%
where $G^\prime(\zeta)$ is the real part. Eqs. \ref{eq:Sigma_fourier_passive_coherent} and \ref{eq:Sigma_fourier_active_coherent}, say that integrating the active spatial filter with the imaging system gives an additional standard deviation $\sigma_{\zeta _q,Aux}$, dependent on the filter parameters. Eq. \ref{eq:Sigma_fourier_aux_coherent} shows how the active filter parameters, such as $A_0$, the center frequency $q_c$, and the width of $G(\zeta)$ contribute to the noise at each Fourier component. Before proceeding further, we reduce Eq. \ref{eq:Sigma_fourier_aux_coherent} to
\begin{equation}
\sigma^2_{\zeta _q,Aux} \approx \frac{1}{M}\mathlarger{\sum \limits ^{(\frac{M}{2}-1)\Delta k_y} _{\zeta=-\frac{M}{2}\Delta k_y}}  |O(\zeta)T(\zeta)  |^{2} A_0^2\sigma _{sd}^2  |G(\zeta)|^2,
\label{eq:Sigma_fourier_aux_coherent_reduced}
\end{equation}since the summation of the first term inside the brackets in Eq. \ref{eq:Sigma_fourier_aux_coherent} can be generally dropped. For example, consider compensating the spatial frequencies  $10k_0\leq \zeta \leq12k_0$. According to the green line in Fig. \ref{fig:Fig_TransferFunction}, the estimated value for $A_0$ is approximately within the order $10^6 \sim 10^8$.

For simplicity, we rewrite the transfer function of the active spatial filter in Eq. \ref{eq:Active_spatial_filter} as
\begin{equation}
A_R(\zeta)= 1 + A_0G_R(\zeta),
\label{eq:Active_spatial_filter1}
\end{equation}
where $G_R(\zeta) = |G_R(\zeta)|\mathrm{e}^{i\varphi_{R} (\zeta)}$ is a unit magnitude rectangular function of width $Wk_0$ and centered at $\zeta_c$. That is
\begin{equation}
|G_R(\zeta)| = \left\{ \begin{array}{cr}
                1 & \hspace{5mm} \bigg |\frac{(\zeta - \zeta_c)}{Wk_0}\bigg | \leq \frac{1}{2} \\
                0 & \hspace{5mm} \texttt{otherwise} .\\
                \end{array} \right .
\label{eq:Rect_function}
\end{equation}
This redefinition conveniently emphasizes the effect of selective spectral amplification without loss of generality. Then, we can express the ratio $R_{\sigma}$ of $\sigma^2_{\zeta _q,Aux}$ in Eq. \ref{eq:Sigma_fourier_aux_coherent_reduced} to $\sigma^2_{\zeta _q,P}$ in Eq. \ref{eq:Sigma_fourier_passive_coherent} as
\begin{equation}
R_{\sigma} =  A_0^2 \frac{P_{I_P,W}}{P_{I_P}} ,
\label{eq:Sigma_fourier_ratio_aux_passive_rect1}
\end{equation}
where $P_{I_P,W}$ is the portion of the total power contained by $I_P(\zeta)$ distributed over bandwidth $Wk_0$ and centered at $\zeta_c$, and $P_{I_P}$ is the total power contained by $I_P(\zeta)$. Thus, it is important to note that Eq. \ref{eq:Sigma_fourier_ratio_aux_passive_rect1} is the ratio of the power in the selectively amplified band to the total power in the noiseless signal without selective amplification. This ratio should not be too large to prevent excessive noise amplification.
Consider, for example, the case where the Fourier components within $6k_0 \leq \zeta \leq 8k_0$ are selected for amplification. We set $A_0 =10^4$ to guarantee the recovery of this entire band. This overcompensates the lower spatial frequencies within the band. From Eq. \ref{eq:Sigma_fourier_ratio_aux_passive_rect1} the ratio $R_{\sigma}$ evaluates to about $13$. This indicates that even though Fourier components were strongly amplified, the resultant increment in the spectral variance is comparatively small. This can be generalized for an arbitrary $\zeta_c$. Fig. \ref{fig:Fig_PSD} shows the power spectral density (PSD) plots for $|O(\zeta)|^2$ and the corresponding $|I_P(\zeta)|^2= |O(\zeta)T(\zeta)|^2$ indicated by black and red lines, respectively. Since $|T(\zeta)|$ decays with increasing $q$ (see black line in Fig. \ref{fig:Fig_TransferFunction}), the PSD of $I_P(\zeta)$ clearly follows a similar trend. Most of the power contained within the image is distributed over a small portion of the Fourier spectrum. For example, if the previously selected band is modified to $8k_0 \leq \zeta \leq 10k_0$, $P_{I_P,W}$  in Eq. \ref{eq:Sigma_fourier_ratio_aux_passive_rect1} will decrease as can be seen from Fig. \ref{fig:Fig_PSD}. However, $P_{I_P}$ will remain the same and therefore, the ratio between $P_{I_P,W}$ and $P_{I_P}$ decreases. This conveniently restricts $R_{\sigma}$ from becoming large even though the Fourier components within $8k_0 \leq \zeta \leq 10k_0$ require larger amplification compared to the previous example.

Based on $R_\sigma$ in Eq. \ref{eq:Sigma_fourier_ratio_aux_passive_rect1} the ACI technique suggests, in principle, an infinite resolution. Because there is a trade-off between the illumination intensity and the bandwidth of the passive spatial filter to keep the spectral SNR above $0\,$dB at an arbitrarily large spatial frequency in the loss compensated image spectrum. Once the SNR is above $0\,$dB for a particular spatial frequency, that particular frequency of the image can be reconstructed with deconvolution using the active transfer function in Eq. \ref{eq:Integrated_transfer_function}. The larger the illumination intensity, the smaller the bandwidth should be to suppress the noise amplification. However, in practice the resolution is limited by several factors: maximum power, minimum bandwidth and maximum center frequency of the passive spatial filter, and the minimum pixel size. Also, the present model of ACI does not consider weak signals, which should be treated with a quantum optical model \cite{Haus}.

\bigskip

\subsection{Improving SNR and resolution limits}
The above inhibition of noise amplification during the compensation process results in substantial improvement in system performance \cite{ghoshroy2017active,Ghoshroy:18,doi:10.1021/acsphotonics.7b01242}. This is investigated by comparing between the spectral SNR of the passive and active systems. A general expression for the spectral SNR is
\begin{equation}
SNR(\zeta) = \frac{|I(\zeta)|}{\sigma _{\zeta _q}}.
\label{eq:SNR_General}
\end{equation}
Substituting the constant $A_0$ with a functional form $A_0(\zeta) = |T(\zeta)|^{-1}$ allows optimal amplification for full compensation of losses within $Wk_0$ bandwidth and is adopted below to emphasize the relative importance of the selective amplification rather than the exact functional form. Alternatively, a Gaussian or log-normal form of $|G_R(\zeta)|$ can also be used to better describe the previously considered MM spatial filters \cite{PhysRevApplied.10.024018,Ghoshroy:18}. Additionally, for the remainder of this work we will use $\sigma_{sd} = 10^{-3}$ in the signal-modulated noise model in Eq. \ref{eq:Signal_modulated_noise_model} for consistency with our previous works \cite{ghoshroy2017active,PhysRevApplied.10.024018,Ghoshroy:18,doi:10.1021/acsphotonics.7b01242}, where an experimental imaging system detector \cite{Akiba2010} is considered.

\begin{figure}[htb]
\centering
\includegraphics[width=0.95\linewidth]{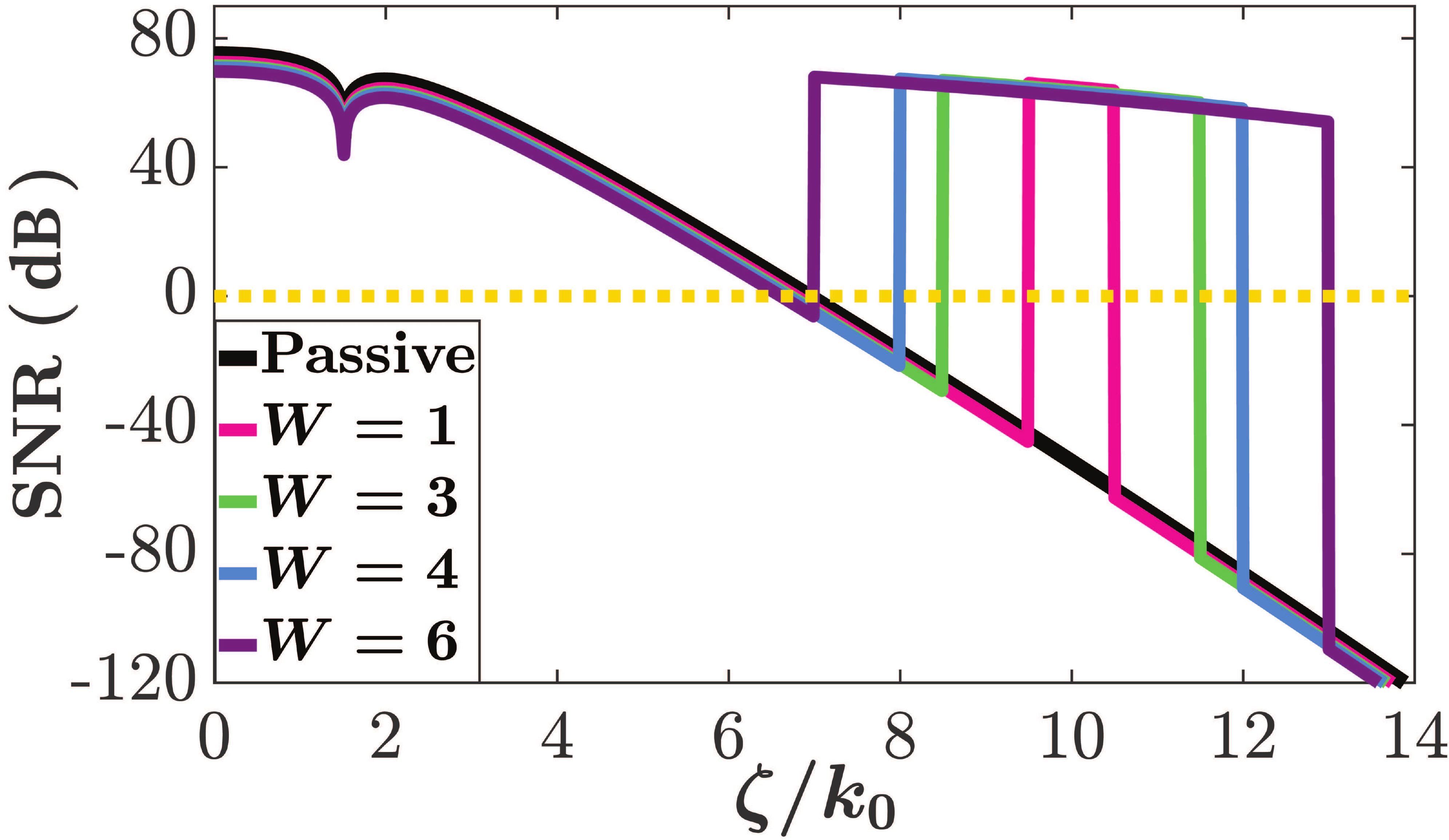}
\caption{SNRs of the passive (black line) and active imaging systems (pink, green, blue and purple lines) with different $W$. The effective resolution limit of the passive imaging system is approximately $\zeta = 7k_0$. In contrast, the SNR of the active imaging systems incorporating ACI is increased within the selected bands of each filter. Slightly reduced SNR outside the selected bands indicate the noise contribution from the auxiliary source.}
\label{fig:Fig_SNR}
\end{figure}

Based on Eqs. \ref{eq:Images_coherent} and \ref{eq:SNR_General}, the SNR of the passive and active imaging systems $SNR_P(\zeta)$ and $SNR_A(\zeta)$, respectively, are written as
\begin{equation}
SNR_P(\zeta) = \frac{|O(\zeta)T(\zeta)|}{\sigma _{\zeta _q,P}},
\label{eq:SNR_Passive}
\end{equation}
and
\begin{equation}
SNR_A(\zeta) = \frac{|O(\zeta)T(\zeta)||1 + A_0(\zeta)G_R(\zeta)|}{\sigma _{\zeta _q,A}},
\label{eq:SNR_Active}
\end{equation}
respectively. $SNR_P(\zeta)$ is plotted by the black line in Fig. \ref{fig:Fig_SNR} and $SNR_A(\zeta)$ for filters with $W=1,3,4$, and $6$ by the pink, green, blue and purple lines, respectively. Note that $\zeta_c$ is kept constant at $10k_0$ and the dashed yellow line marks $SNR=0\,$dB. The intersection of $SNR_P(\zeta)$ with the dashed line marks the resolution limit of the passive system since larger Fourier components will be indistinguishable from noise in the detected signal. However, $SNR_A(\zeta)$ shows a remarkable improvement especially within the regions where compensation is provided. We point out that $SNR_A(\zeta)$ is less than $SNR_P(\zeta)$ outside the selected bands as expected, since the additional noise from $\sigma_{\zeta _q,Aux}$ affects the entire spectrum (see Eq. \ref{eq:Sigma_final_fourier}). This contribution increases with $W$ as is evident from Fig. \ref{fig:Fig_SNR}. Nevertheless, the additional increment in the spectral variance is significantly smaller than the amplification provided to each Fourier components inside the selected bands, which results in an impressive enhancement in SNR. The purple line is particularly interesting since it encapsulates the remarkable power of ACI. The rectangle function $|G_R(\zeta)|$, in this case, spans a fairly broad $6k_0$ bandwidth and has essentially extended the resolution limit of the system close to double compared to the passive system.

\subsection{Arbitrary objects} 
\begin{figure}[htb]
\centering
\includegraphics[width=0.95\linewidth]{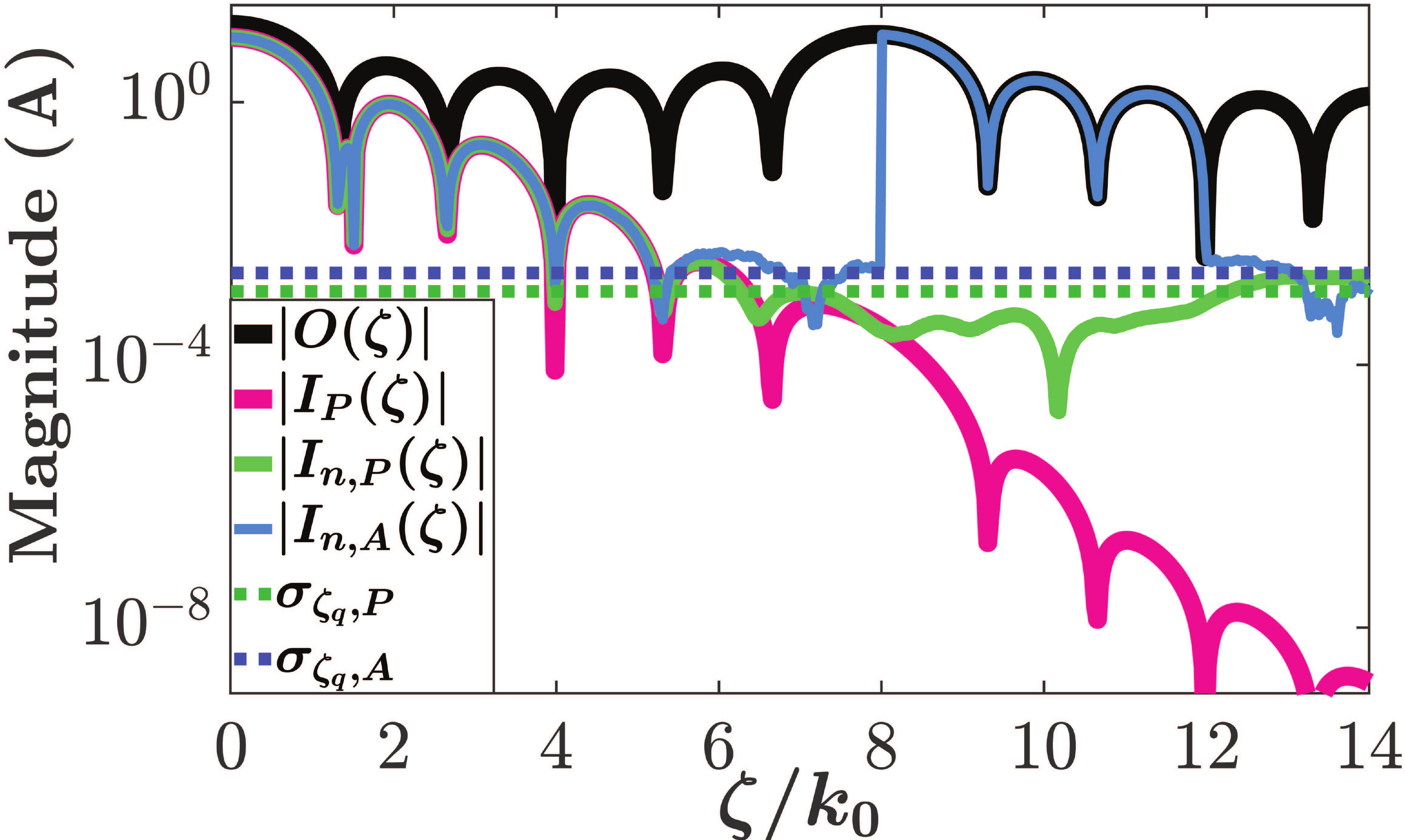}
\caption{Generalization of the ACI to an arbitrary object. The amplitude of the Fourier transforms of the object, the corresponding noise-free and noisy passive images, and the active image  are shown by the black, pink, light green and light blue lines, respectively. The standard deviations for the noisy passive image and the active image spectra are shown by the dashed dark green and dark blue lines, respectively. The active image is adequately compensated within $8k_0 \leq \zeta \leq 12k_0$ with a very small amplification of SD noise.}
\label{fig:Fig_Coherent}
\end{figure}

In general, the theory of ACI can be expanded to arbitrary objects. This is illustrated with Fig. \ref{fig:Fig_Coherent} where the Fourier spectrum of an arbitrary object is plotted by the black line. The corresponding noise-free passive image spectrum is calculated from Eq. \ref{eq:Images_coherent_passive} and corrupted with noise in the spatial domain according to the signal-modulated noise model in Eq. \ref{eq:Signal_modulated_noise_model}. The noisy image is then Fourier transformed to obtain $I_{n,P}(\zeta)$. The magnitudes of $I_{P}(\zeta)$ and $I_{n,P}(\zeta)$ are shown in Fig. \ref{fig:Fig_Coherent} by pink and light green lines, respectively. The standard deviation $\sigma_{\zeta_q,P}$, which has degraded the passive image spectrum is shown by the dashed dark green line. We can see how $|I_P(\zeta)|$ progressively worsens  with increasing $\zeta$. Eventually, $\sigma_{\zeta_q,P}$ becomes comparable to $|I_{P}(\zeta)|$ at approximately $\zeta = 7k_0$ after which $|I_{n,P}(\zeta)|$ is overwhelmed by noise, similar to the simpler Gaussian object in Fig. \ref{fig:Fig_SNR} (see black line). The noise-free active image spectrum is calculated from Eq. \ref{eq:Images_coherent_active} taking $\zeta_c=10k_0$, $W=4$, and substituting $A_0G(\zeta)$ with $ A_0(\zeta)G_R(\zeta)$, where $A_0(\zeta) = |T(\zeta)|^{-1}$. This active image is then also corrupted with noise in the spatial domain and Fourier transformed to obtain $I_{n,A}(\zeta)$, magnitude of which is shown by the light blue line in Fig. \ref{fig:Fig_Coherent}. The standard deviation for the active system $\sigma_{\zeta _q,A}$ is shown by the dashed dark blue line.

Fig. \ref{fig:Fig_Coherent} clearly manifests the noise-resistant effect of the selective amplification. Note that the missing nodes on the object spectrum are accurately recovered inside the band $8k_0 \leq \zeta \leq 12k_0$ where the selective amplification is provided. The inhibition of noise amplification with ACI's selective spectral amplification is therefore applicable for arbitrary objects. Based on Fig. \ref{fig:Fig_Coherent}, we find that the resolution limit can be in the end extended by more than $40\,\%$ when the underlying passive spatial filter is illuminated with an intensity of about $0.03\mathrm{mW/\upmu m^2}$ (i.e., the intensity amplification factor $A_0^2\approx 1.6 \times 10^5$). A coherent light around this level of intensity is accessible for a superresolution imaging experiment with a relatively less lossy MM structure \cite{liu2020personal}. Retrieving deep subwavelength information from a low-Q system disturbs the system and lets the high spatial frequency modes quickly dissipate. In our calculations we did not include this effect. Therefore, it is necessary to maintain a sufficiently high intensity continuous wave illumination to counter this effect. Here, for simplicity, we consider only one-dimensional imaging. Therefore, the dimension of the pixel along the $z$-direction can be taken equal to the length $L$ of the image plane. For two-dimensional imaging more intensity is needed. Additional noise and speckle associated with high illumination intensity and small pixel size limit the achievable resolution. Also, the amplification at high spatial frequencies will be difficult in the presence of spatial dispersion. Therefore, the designed spatial filter in Fig. \ref{fig:Fig_Schematic}(b) should support the highest desired spatial frequency. In a future work, an efficient (i.e., low power) implementation could replace the spatial filter in Fig. \ref{fig:Fig_Schematic}(b) with a plasmonic structured illumination \cite{bezryadina2018high} that is systematically designed based on Eqs. \ref{eq:Sigma_final1} and \ref{eq:Sigma_final_fourier}. Another possibility could be considered along the lines of \cite{xiong2007two}, where the HMM is used to collect the low and high spatial frequencies at different polarizations. In our model, we assume $60\,$dB spatial SNR (i.e., $\sigma_{sd}=10^{-3}$). Since the spectral variance is constant through the image spectrum (see Fig. \ref{fig:Fig_Coherent}) and proportional to $\sigma_{sd}^2$, the larger the spatial SNR the lower the required power is to reconstruct a specific spatial frequency.
\begin{figure}[htb]
\centering
\includegraphics[width=0.95\linewidth]{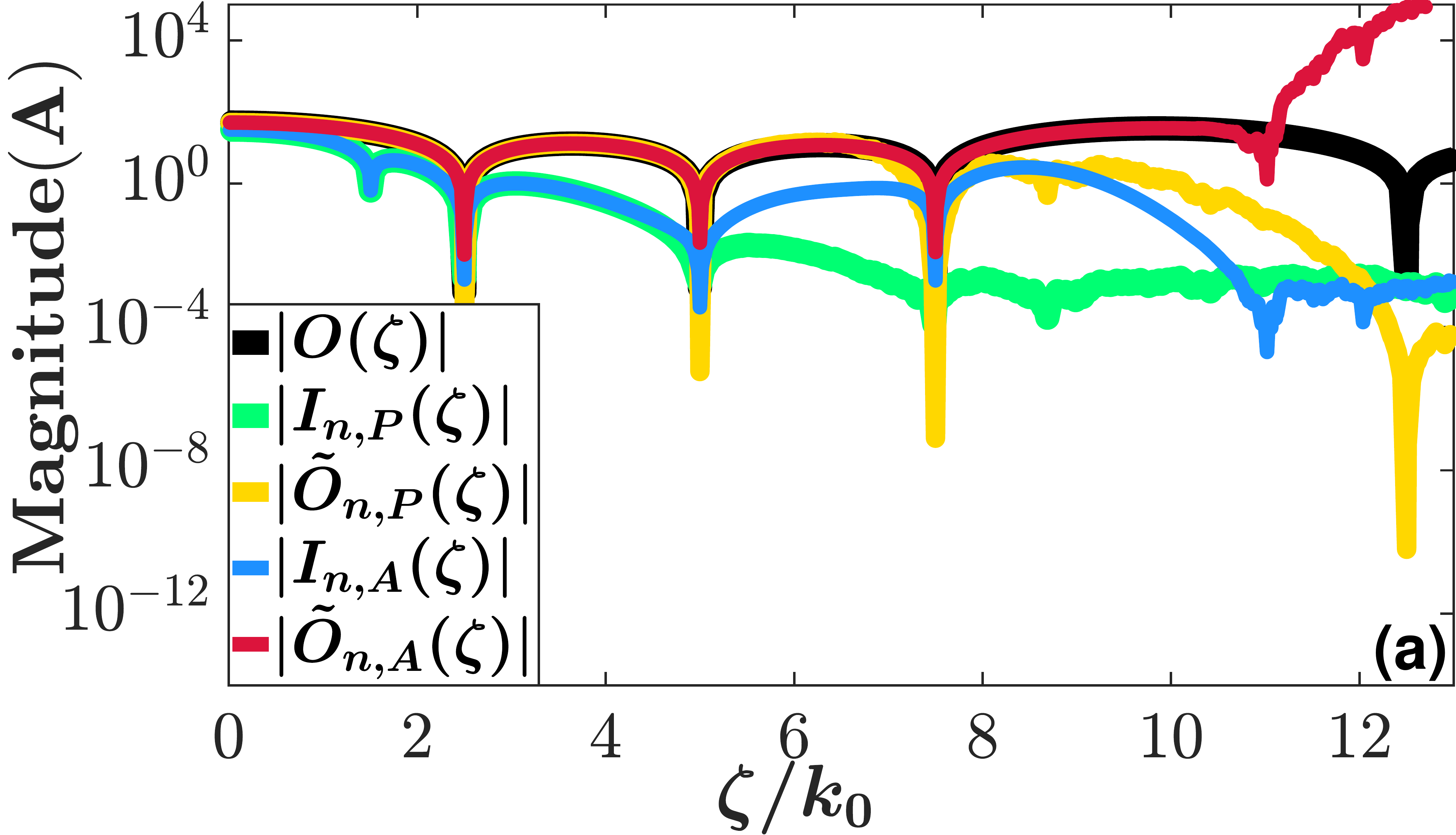}\\
\includegraphics[width=0.95\linewidth]{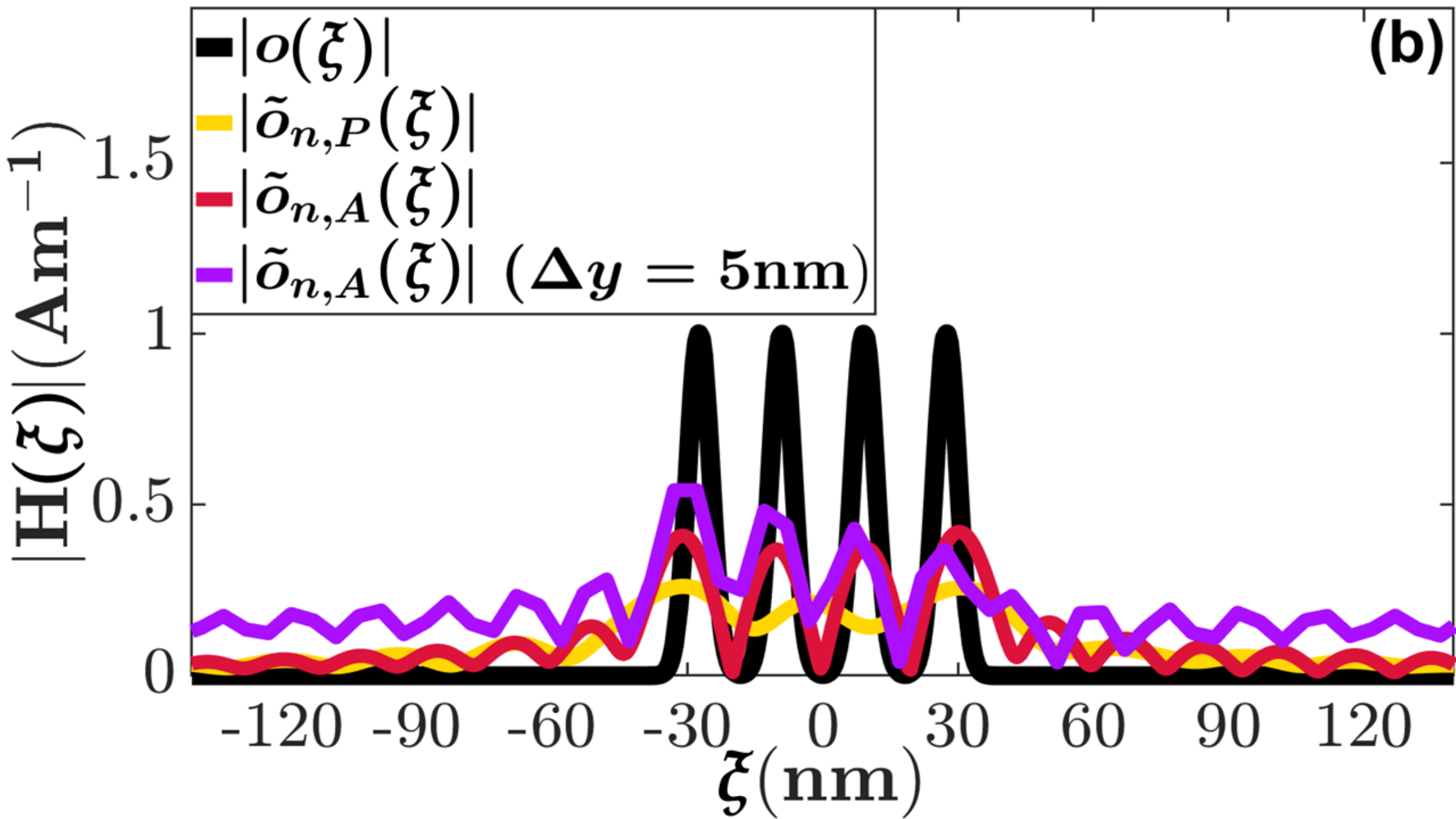}
\caption{The comparison of the image reconstruction using ACI (without optimal Wiener filter) and the passive image reconstruction using the optimal Wiener filter, in (a) Fourier domain and (b) spatial domain. $\tilde{O}_{n,P}(\zeta)$ and $\tilde{O}_{n,A}(\zeta)$ refer to passive reconstructed image spectrum obtained by the optimal Wiener filter and the active reconstructed image spectrum obtained by the ACI without optimal Wiener filter. $\tilde{o}_{n,P}(\xi)$ and $\tilde{o}_{n,A}(\xi)$ are the respective reconstructed images in the spatial domain. The object (black) and the noisy images $\tilde{I}_{n,P}(\zeta)$ and $\tilde{I}_{n,A}(\zeta)$ before deconvolution are also indicated. Only magnitudes are shown. In (b) the discrete space $\xi$ is interpolated to guide the eye. The ACI method clearly resolves the 4 objects (see black lines) separated by $18\,$nm peak-to-peak distance, using 2 overlapping Gaussian pass-bands with FWHMs of $2.3k_0$ and centered at $7k_0$ and $8.5k_0$. The constant amplitude amplification factor $A_0=630$.}
\label{fig:Reconstruction}
\end{figure}

In the ACI method, once the selective amplification is applied to the spatial frequencies that were previously buried under the noise (see Fig. \ref{fig:Fig_Coherent}), the reconstructed image can be obtained from deconvolution based on the active transfer function (see Eq. \ref{eq:Integrated_transfer_function}). If the optimal Wiener filter $[1 + 1/SNR_A(\zeta)^2]^{-1}$ \cite{roggemann1992linear,biemond1990iterative,zaknich2005principles} is used in the deconvolution step, the reconstructed image can be written as
\begin{align}
\tilde{O}_{n,A}(\zeta) &= I_{n,A}(\zeta) \frac{1}{1 + \frac{1}{{SNR_A(\zeta)}^2}} \notag\\
& \times\frac{1}{T(\zeta)[1 + A_0G(\zeta)]},
\label{eq:SNR_Passive}
\end{align}
assuming a constant $A_0$ and general pass-band function $G(\zeta)$. For high $SNR_A(\zeta)$ (i.e., around low spatial frequencies and regions of selective amplification), this deconvolution process approaches to ``active inverse filtering.'' For low $SNR_A(\zeta)$ (e.g., around $7k_0$ and $13k_0$ in Fig. \ref{fig:Fig_Coherent}), the optimal Wiener filter does not heavily amplify the noise as opposed to the inverse filter \cite{roggemann1992linear}. In general, the optimal Wiener filter employs the image SNR (see Eq. \ref{eq:SNR_Passive}) to prevent excessive noise amplification \cite{biemond1990iterative,zaknich2005principles}. In this regard, the selective amplification in the ACI method has also a similar spirit as the optimal Wiener filter. However, as shown below in Fig. \ref{fig:Reconstruction}, with the ACI even without the optimal Wiener filter one can restore spatial frequencies that cannot be restored by a typical optimal Wiener deconvolution \cite{roggemann1992linear,biemond1990iterative,zaknich2005principles}. This is achieved by selectively amplifying those spatial frequencies, while preventing excessive noise amplification in accordance with Eqs. \ref{eq:Sigma_final1} and \ref{eq:Sigma_final_fourier}. Furthermore, as given in Eq. \ref{eq:SNR_Passive}, when the optimal Wiener filter is integrated with the ACI, this extends the restored spatial frequency range of the optimal Wiener filter.

Fig. \ref{fig:Reconstruction} compares the passive and active reconstructed images and their corresponding spectra for 4 Gaussian objects separated by $18\,$nm peak-to-peak distance. The yellow line corresponds to the unresolved passive image reconstructed with the optimal Wiener filter. The images reconstructed with the ACI are shown with the red and purple lines in Fig. \ref{fig:Reconstruction}(b). The purple line highlights the effect of discretization assuming $\Delta y =5\,$nm. Consistent with Eq. \ref{eq:Sigma_final_fourier}, the noise is increased with larger discretization. It is clearly seen that the objects are fully resolved using ACI, with a resolution better than $\lambda /20$. The selective amplification process of ACI is achieved by 2 overlapping Gaussian pass-bands with FWHMs of $2.3k_0$ and centered at $7k_0$ (i.e., near the resolution limit of the passive system) and $8.5k_0$. The incident plane wave illumination amplitude [see Fig. \ref{fig:Fig_Schematic}(b)] is increased by a constant factor of $A_0=630$. Finally, the reconstructed images are obtained from deconvolution based on the active transfer function (see Eq. \ref{eq:Integrated_transfer_function}) and using the noisy active image spectrum $I_{n,A}(\zeta)$ [see blue line in Fig. \ref{fig:Reconstruction}(a)].

\section{Discussion and Conclusion}
The ACI is more than a loss compensation in MMs or plasmonics. The ACI concept, the then-called $\Pi$ scheme, was first numerically demonstrated as a loss compensation method in a plasmonic NIM \cite{sadatgol2015plasmon}, but later rapidly evolved into a scheme for the mitigation of information loss in noisy and lossy linear systems. The ACI has, since, turned into a scheme for spectrum manipulation using selective amplification and correlations \cite{ghoshroy2017active,Ghoshroy:18,doi:10.1021/acsphotonics.7b01242}.

In this work, we have presented a mathematical analysis of the conceptual framework of ACI. We showed that selective amplification of a controllable band of spatial frequencies with an auxiliary source can provide sufficient amplification to previously attenuated spatial frequencies with minimal amplification of noise. It is important to emphasize that the amplification process in the theory of ACI described here is fundamentally different than the traditional optical gain media and does not require a quantum optical model. The ACI is more feasible than optical gain or nonlinear media, which are more complex and cumbersome due to pumping, gain saturation, or amplified spontaneous emission (ASE) \cite{prasad1994implications,Haus}. In the coherent model of ACI, the amplification of the spatial frequencies within the selected band (see Figs. \ref{fig:Fig_SNR} and \ref{fig:Fig_Coherent}) is achieved by the coherent superposition of the original object field with an external auxiliary source, which is correlated with the object field (see Eq. \ref{eq:Images_coherent}). Possible physical generations of the auxiliary source relying on the HMMs and injection of plasmons have been studied in detail in our previous works \cite{sadatgol2015plasmon,PhysRevApplied.10.024018,Ghoshroy:18}. The implementations for far-field imaging can be made possible with, for example, structured illumination \cite{ingerman2018signal} and spatial filtering \cite{becker2018better,adams2019enhancing}. Thus, the ACI does not suffer from the severe adverse effect of ASE on the SNR associated with the amplification of weak signals using optical gain media \cite{prasad1994implications,Haus}. Also, the imaging system here employs amplification (e.g., by using a brighter source) prior to the lossy transmission to avoid the difficulty with the signal amplification at the detection side, especially for the retrieval of higher spatial frequencies, and is operated with stronger signals. Moreover, the classical correlations play an important role in ACI \cite{ghoshroy2017active,qian2017emerging}. In practice, the ACI may not necessarily need increased input power or a separate auxiliary source, but may only need to locally (selectively) amplify the signal spectrum by redistributing the spatial frequency content \cite{becker2018better,adams2019enhancing}.

The present model of ACI is based on linear systems. Therefore, ACI can also compensate the adverse effect of nonlocality \cite{demetriadou2008taming} on the imaging performance of the system, as long as the nonlocal system operates in the linear regime. However, since such a nonlocal system has a poor transfer function compared to the one without spatial dispersion, it will be more difficult to extend the resolution limit. On a similar token, we have previously shown in \cite{Zhang:17} that an adverse effect of a deviation from homogeneous effective medium approximation can also be compensated with ACI.

We provided a detailed analytical explanation of the role and importance of the various  aspects of ACI for greater insights into the previous results \cite{Ghoshroy:18,ghoshroy2017active}. The same mathematical framework can be further expanded to include incoherent illumination \cite{doi:10.1021/acsphotonics.7b01242,ingerman2018signal,becker2018better,adams2019enhancing} using the Wiener-Khinchin theorem. We believe that this work can also  theoretically explain the other numerical and experimental results presented in  independent works including pattern uniformity in lithography \cite{liang2018achieving}, high-resolution Bessel beam generation \cite{liu2017nanofocusing}, and acoustic real-time subwavelength edge detection \cite{moleron2015acoustic}, and fosters further explanation of recent simulation and experimental results in far-field imaging \cite{ingerman2018signal,becker2018better}.

Eqs. \ref{eq:Sigma_final1} and \ref{eq:Sigma_final_fourier} can also be used to explain why dark-field imaging \cite{benisty2012dark,repan2015dark,shen2017hyperbolic} improves the contrast. Blocking the low spatial frequencies reduces the power contained in the signal, hence the flat variance in the Fourier spectrum. Because the high spatial frequencies are not blocked, however, the spectral SNR increases in this region, hence the image contrast. In contrast, as can be seen from the active transfer function in Eq. \ref{eq:Integrated_transfer_function}, the ACI does not aim to block low spatial frequencies. It rather aims to selectively amplify the high spatial frequencies buried under the noise without excessive noise amplification. Therefore, it does not sacrifice brightness or strongly enhance artifacts unlike dark-field imaging \cite{repan2015dark}. The theory presented here can also be applied to bright field imaging. Guided by Eqs. \ref{eq:Sigma_final1} and \ref{eq:Sigma_final_fourier}, the high spatial frequencies can be recovered with the selective amplification.

Revealed from the simple mathematical result in Eq. \ref{eq:Sigma_final_fourier}, we conjecture that the theoretical concepts of ACI can be potentially generalized to numerous scenarios in noisy and lossy linear systems (e.g., atmospheric imaging \cite{hanafy2014detailed,hanafy2015estimating,hanafy2015reconstruction,ghoshroy2019super}, bioimaging \cite{Vadivambal}, deep-learning based imaging \cite{wang2019deep}, structured illumination \cite{ingerman2018signal}, tomography \cite{guillet2014review}, time-domain spectroscopy \cite{guerboukha1018toward,ahi2019a}, free space optical communications \cite{gbur2002spreading,dogariu2003propagation,gbur2014partially,hyde2018controlling}, ${\cal PT}$ symmetric non-Hermitian photonics \cite{monticone2016parity,ganainy2019dawn,li2019virtual}, and quantum computing \cite{gueddana2019can,gueddana2019toward,ghoshroy2019super}, etc.) at different frequencies. Since ACI operates down at the physical layer, all of these scenarios should benefit from ACI for improved performance. Analogous equations to Eqs. \ref{eq:Sigma_final1} and \ref{eq:Sigma_final_fourier} can be derived for different systems to understand the noise behavior and other effects (e.g., turbulence, scattering, aberration, dispersion, etc.) in the output spectrum to determine the best amplification strategy.

\bigskip
\noindent
\textbf{Funding.} Office of Naval Research (award N00014-15-1-2684).

\noindent
\textbf{Acknowledgments.} The authors would like to thank Jeremy Bos at Michigan Technological University for fruitful discussions and Jasmine O'Hanlon-Mundy for graphics.

\noindent
\textbf{Disclosures.} The authors declare no conflicts of interest.



\end{document}